\documentclass{article}
\usepackage{amsfonts}
\usepackage[dvips]{graphicx}

\begin{document}

\title{Vector field theories in cosmology}
\author{A. Tartaglia and N. Radicella \\
Dipartimento di Fisica del Politecnico and INFN section of Turin\\
Corso Duca degli Abruzzi 24, I-10129 Torino, Italy\\
e-mail: angelo.tartaglia@polito.it and ninfa.radicella@polito.it}
\maketitle

\begin{abstract}
Recently proposed theories based on the cosmic presence of a vectorial field
are compared and contrasted. In particular the so called Einstein aether
theory is discussed in parallel with a recent proposal of a strained
space-time theory (Cosmic Defect theory). We show that the latter fits
reasonably well the cosmic observed data with only one, or at most two,
adjustable parameters, whilst other vector theories use much more. The
Newtonian limits are also compared. Finally we show that the CD theory may
be considered as a special case of the aether theories, corresponding to a
more compact and consistent paradigm.
\end{abstract}

\section{Introduction}

The most successful field theories of the XXth century are in general tensor
theories on a four-dimensional manifold. This is true for the
electromagnetic field as well of course as for the gravitational
interaction. In the former case the "root" of the theory is in a four-vector
potential, in the latter also the potential is a rank 2 symmetric tensor
(the metric tensor).

Although, properly speaking, "tensor" includes any rank from $0$ (a scalar)
up to higher values, by "tensor" theory one normally means a theory based on
an at least rank 2 tensor. On this respect we shall here discuss "vector"
theories as rank 1 tensor theories.

Theories of this sort have not been considered as frequently as scalar and
scalar-tensor theories. Of course we need a motivation and ours has been at
the cosmic scale. Since 1998 evidence has been found pointing at an
accelerated expansion of the universe (see \cite{acceleration}) and
theorists have been working since then to find an explanation of the alleged
behaviour. In various forms a sort of dark energy fluid has been envisaged
(see \cite{darkenergy1} and \cite{darkenergy2}) or modifications of the
classical General Relativity theory have been proposed producing the sought
for effects (see \cite{mod1} and \cite{mod2}). One of us has put forth the
idea that a cosmological vector field be responsible for the acceleration:
let us call the one based on this vector, "Cosmic Defect" (CD) theory \cite%
{dark1}.

Another cosmic vector field has been discussed in the literature in recent
years in a group of theories that we shall call here, for short, Vector \AE %
ther Theories (VET), although their authors have used different names or no
name at all. One of these theories has indeed been christened Einstein \AE %
ther (\AE ) \cite{EA}. In all cases by "\ae ther" the cosmic vector field is
meant.\textbf{\ }

The initial motivation for the VET was not related to cosmology, but rather
to fundamental quantum field theory, where reasons exist to doubt of an
exact Lorentz invariance. In fact, the Lorentz group is non-compact
and leads to divergences in quantum field theory, associated with states of
arbitrary high energy and momentum. Furthermore, because of the
non-compactness of the group, it is not possible to experimentally test the
invariance at all scales of energy.

Considering a $D$-dimensional universe with $D - 4$ compactified
extra-dimen\-sions, the so-called tensor-induced Lorentz symmetry breaking has
been investigated in \cite{kos sam}, and a general discussion of the purely
gravitational and cosmological aspects has been made.

Many other approaches to field theory, like non-commutative field
theory \cite{carroll et} as well as non-string approaches to quantum gravity
( see references in \cite{kos sam}) suggest Lorentz violation.

In practice the presence of a vector field in a four dimensional
space-time selects locally a special direction (in \cite{kos sam}
also the possible breaking in the extra-dimensions is discussed). If the
vector is time-like the local Lorentz symmetry is somehow broken and this
has been the starting point for VET \cite{EE1}. Of course, once the global
existence of a vector field has been hypothesized, its presence will bring
about consequences also at the cosmic scale and here is where an
intersection with the CD theory is found.

Actually, since the '70's a vector theory of gravity was proposed,
somehow generalizing the Brans-Dicke theory, in which gravitation was
produced by a rank-2 tensor field and a scalar field. The authors in \cite{will} looked for detectable effects due to the motion of the solar system
relative to a preferential reference frame and discussed both cosmological
and solar system constraints.

The local Lorentz symmetry breaking should however not be overestimated. A
cosmic privileged reference frame actually exists: the one of the cosmic
microwave background radiation (CMB) or, which is the same, in any Robertson
Walker (RW) universe (as ours is commonly assumed to be, on the average) the
co-moving frame of the cosmic "fluid" of galaxies. This local selection of a
privileged time axis, or the equivalent Lorentz symmetry breaking, shows up
only on cosmic time scales when the effects of the expansion appear.
Locally, in usual time scales nothing is manifested even though the
"privileged" frame is there, unless possibly for phenomena at very high
energies.

Leaving these comments on the background the points on which one would like
to gauge a theory are:

- ability to reproduce a given observed behaviour;

- absence of unwanted "side effects";

- simplicity;

- existence of a reasonable interpretation frame or paradigm for the theory
to fit in.

On these points we shall elaborate throughout the paper. In Sec. \ref{sec2}
we shall review the VE theories, inspecting the Lagrangian and deriving the
equations of motion; in Sec. \ref{sec3} the same will be done with the CD
theory. In the following two sections the results of both the VET
and the CD theory will be analyzed in two respects: in Sec. \ref{sec4} the cosmological solution will be derived, looking for an accelerated
expansion; in Sec. \ref{sec5} the Newtonian limit is discussed and in both
cases the Poisson equation is recovered with a rescaling of the coupling
constant. In Sec. \ref{sec6} we shall show that the CD theory may
be viewed as a special case of the VET and the Standard Model
Expansion (SME) theory, and finally we shall draw some conclusions (Sec. %
\ref{conc}).

\section{Vector \AE ther Theories}

\label{sec2}

Adopting the traditional approach to field theory, we may write the total
action integral for a space time containing a vector field $U$ as the sum of
three parts

\begin{equation}
S=\int \frac{d^{4}x}{c}\sqrt{-g}\left( \frac{1}{2\kappa }R+\mathcal{L}_{U}+\mathcal{L}%
_{m}\right) .  \label{zero}
\end{equation}%
Of course $g$ is the determinant of the metric tensor and $R$ is the scalar
curvature of the manifold. $\mathcal{L}_{U}$ is the Lagrangian density of
the vector field, and $\mathcal{L}_{m}$ is the one of matter; $\kappa =\frac{%
8\pi G}{c^{4}}$ is the coupling constant between matter and geometry.
Writing the action in the form (\ref{zero}) we are implicitly
assuming that no direct coupling between the vector field $U$ and
matter exists; both the vector field and matter couple with geometry.

According to \cite{ferreira} and \cite{zlos1}, the most general form for the
\AE ther Lagrangian density in the action integral (\ref{zero}) for a
vector-tensor theory, including terms up to second order in derivatives and
fourth order in the fields, is written as follows:

\begin{equation}
\mathcal{L}_{U}=K_{\mu \nu }^{\alpha \beta }\nabla _{\alpha }U^{\mu }\nabla
_{\beta }U^{\nu }-k(U^{\mu }U_{\mu }-M^{2}n^{\alpha }n_{\alpha })^{2},
\label{fergri}
\end{equation}%
We can recognize a kinetic and a potential term for the vector field. The
coefficients of the kinetic term are contained in the rank 4 tensor:
\begin{eqnarray}
K_{\mu \nu }^{\alpha \beta } &=&K(1)_{\mu \nu }^{\alpha \beta }+K(2)_{\mu
\nu }^{\alpha \beta },  \nonumber \\
K(1)_{\mu \nu }^{\alpha \beta } &=&c_{1}g^{\alpha \beta }g_{\mu \nu
}+c_{2}\delta _{\mu }^{\alpha }\delta _{\nu }^{\beta }+c_{3}\delta _{\nu
}^{\alpha }\delta _{\mu }^{\beta }+c_{4}U^{\alpha }U^{\beta }g_{\mu \nu },
\label{Ikappa} \\
K(2)_{\mu \nu }^{\alpha \beta } &=&c_{5}\delta _{\nu }^{\alpha }U^{\beta
}U_{\mu }+c_{6}g^{\alpha \beta }U_{\mu }U_{\nu }+c_{7}\delta _{\mu }^{\alpha
}U^{\beta }U_{\nu }+c_{8}U^{\alpha }U^{\beta }U_{\mu }U_{\nu },  \nonumber
\end{eqnarray}%
in which the terms in $c_{4}$ and $c_{8}$ represent directional covariant
derivatives along the field $U^{\mu }$.

The potential term in (\ref{fergri}) is the gravitational analogue of the
Higgs mechanism of gauge theories, so that the vector field acquires a
vacuum expectation value $Mn^{\alpha }$ that breaks the Lorentz invariance; $%
n$ is a unit four-vector. The action integral written with (\ref{fergri}) is
slightly different from the one that appears in \cite{ferreira} and \cite%
{gripaios} because of the different choice made for constants. Assuming $U$
to be dimensionless, one is left with dimensionless $c_{i}$'s and $M$, too.
Notice, as already stressed before, that the matter Lagrangian
couples only with the metric and not directly with the $U^{\mu }$'s.

The claim of generality on (\ref{fergri}) must be taken with some
caution, because it depends on a number of limiting assumptions on
space-time. Within a wider framework and in the
attempt of constructing a theory in which both General Relativity
and the Standard Model are taken into account, Konsteleck\' y has developed SME (Standard Model Extension), a
theory whose effective Lagrangian contains the fields of the Standard Model
as well as gravity together with additional Lorentz symmetry-violating
terms. The most general formulation of SME \cite{kos} uses an
Einstein-Cartan background, including torsion; in this framework (\ref{fergri}) appears as a special subclass of Lagrangians. An earlier version
of the SME in a Minkowski spacetime had already been studied in \cite
{colladay}.

Konsteleck\'y and Samuel in \cite{kos
sam}, as early as in 1989, considered a Lagrangian density which now could
be seen as a subclass of the VE\ theories. It has been put forth again by
Jacobson and Mattingly \cite{EE1} in 2001, in practice
considering the only K(1) term of (\ref{Ikappa}) and replacing the
potential term by a constraint on the norm of the vector field, introduced
by means of a Lagrange multiplier $\lambda $. The Lagrangian density for the
vector field is then\footnote{%
We shall use a $+---$ signature throughout the paper.}

\begin{equation}  \label{EE}
\mathcal{L}_{U}=K(1)^{\alpha\beta}_{\mu\nu}\nabla_\alpha U^\mu \nabla_\beta
U^\nu + \lambda (U^\alpha U_\alpha-1).
\end{equation}

From now on we shall refer to this theory as Einstein \AE ther (\AE ), from
the name used in \cite{EE1}.

The case analyzed in the earlier formulation corresponds to
choosing all parameters to be zero except for  $c_{1} $and $c_{3}$, for which $ c_{1}+c_{3}=0$ holds.

A theory equivalent to this choice for the \AE ther Lagrangian
had already been studied by Nambu \cite{nambu} in the case of
Minkowski spacetime, who proved it to be equivalent to electrodynamics
in a non-linear gauge. Other contributions in the non-flat background case
are found in \cite{EE1} and \cite{clayton}.

A variant of \AE \cite{Zlosnik} introduces the vector field in the action
in the form of a function $\mathcal{F}$ of the scalar $\mathcal{K}$ obtained
from the $K\left( 1\right) _{\mu \nu }^{\alpha \beta }$ after choosing $%
c_{4}=0$:
\begin{eqnarray*}
\mathcal{K} &=&M^{-2}K_{\mu \nu }^{\alpha \beta }\nabla _{\alpha }U^{\mu
}\nabla _{\beta }U^{\nu } \\
K_{\mu \nu }^{\alpha \beta } &=&c_{1}g^{\alpha \beta }g_{\mu \nu
}+c_{2}\delta _{\mu }^{\alpha }\delta _{\nu }^{\beta }+c_{3}\delta _{\nu
}^{\alpha }\delta _{\mu }^{\beta }.
\end{eqnarray*}

This approach was motivated by its authors, Zlosnik, Ferreira, and~Starkman
(ZFS, for short) by the quest of a modified Newtonian gravity at galactic
scales, as an alternative to dark matter. The Lagrangian for the \AE ther is
now written (\cite{Zlosnik})

\begin{equation}
\mathcal{L}_{U}=M^{2}\mathcal{F}(\mathcal{K})+\lambda (U^{\mu }U_{\mu }-1)
\label{Star}
\end{equation}%
assuming that the Lagrange multiplier $\lambda $ has the dimension of the
inverse of a squared length. Of course (\ref{Star}) coincides with the \AE\ %
Lagrangian when $\mathcal{F}(\mathcal{K})\equiv \mathcal{K}$.

In the general Lagrange density (\ref{fergri}), $U^{\mu }$ is neither
restricted to have a fixed norm nor to be timelike. In the case of a
homogeneous and isotropic universe\footnote{%
The line element is of the form $ds^{2}=c^{2}dt^{2}-a(t)^{2}\delta
_{ij}dx^{i}dx^{j}$.}, however, the assumed space isotropy implies the vector
field to be timelike
\[
U^{\mu }=(U(t),0,0,0),
\]%
and still leaves six free parameters in the equations of motion \cite%
{ferreira}: $c_{2}$, $c_{3}$, $c_{4}$, $c_{7}$, $c_{8}$, and $k$.

This freedom is somewhat reduced in \cite{EA},\cite{EE1}, \cite{Zlosnik} and
in the analysis made by Carroll \cite{Carroll} and Lim \cite{Lim}. All these
authors constrain the vector field to be a unit vector
\[
U^{\mu }=(1,0,0,0),
\]%
and maintain four (see \cite{EA}\cite{EE1}\cite{ea4}), or three (see \cite%
{Zlosnik}\cite{Carroll}\cite{Lim}) free parameters, like in (\ref{EE}) and (%
\ref{Star}).

Varying (\ref{zero}) with respect to the metric tensor elements, we obtain,
as usual, the Einstein equations in the form

\begin{equation}  \label{ein}
G_{\alpha \beta }=\kappa \left(T_{\alpha \beta }^{U}+ T_{\alpha \beta
}^{m}\right),
\end{equation}%
where $T_{\alpha \beta }^{m}$ is the stress-energy tensor for matter, while $%
T_{\alpha \beta }^{U}$ is the one of the vector field. In the case of the
\AE\ theory, the explicit form of $T_{\alpha \beta }^{U}$ is

\begin{eqnarray}
T_{\alpha \beta }^{U} &=&\frac{1}{2}\nabla _{\sigma }\left[ \mathcal{F}%
^{\prime }(J_{(\alpha }^{\,\sigma }U_{\beta )}-J_{\;\alpha }^{\sigma
}U_{\beta )}-J_{(\alpha \beta )}U^{\sigma })\right]  \label{emt} \\
&+&c_{1}\mathcal{F}^{\prime }[(\nabla _{\nu }U_{\alpha })(\nabla ^{\nu
}U_{\beta })-(\nabla _{\alpha }U_{\nu })(\nabla _{\beta }U^{\nu })]+\frac{1}{%
2}g_{\alpha \beta }M^{2}\mathcal{F}+\lambda U_{\alpha }U_{\beta },  \nonumber
\end{eqnarray}%
where
\begin{eqnarray}
\mathcal{F}^{\prime } &=&\frac{d\mathcal{F}}{d\mathcal{K}}  \nonumber \\
J_{\,\sigma }^{\alpha } &=&(K_{\;\;\sigma \gamma }^{\alpha \beta
}+K_{\;\;\gamma \sigma }^{\beta \alpha })\nabla _{\beta }U^{\gamma }.
\nonumber
\end{eqnarray}%
Varying the action with respect to $U^{\mu }$, under the same hypotheses,
one obtains the equations of motion for the vector field

\begin{equation}  \label{vector}
\nabla_\alpha(\mathcal{F}^{\prime \alpha}_{\,\beta})=2 \lambda U_\beta.
\end{equation}


\section{The Cosmic Defect theory}

\label{sec3}

The CD theory, as the theories mentioned in the previous section, ascribes
the behaviour of the universe as a whole to the presence of a cosmic
four-vector field. The difference with respect to the VE\ theories is mostly
in the motivation and interpretation of the vector, then in the choice of
the initial Lagrangian density for space time.

The CD vector field is interpreted in terms of a paradigm considering the
space time as a continuum hosting a defect. The presence of defects (in the
sense of Volterra \cite{volterra}) in a medium implies a permanent state of
stress and strain even in the absence of applied forces from outside. If we
think of the Robertson Walker symmetry, assumed to be correct for describing
the universe, we find an initial singular state giving rise to the symmetry.
In four dimensions we may think as if we had a pointlike defect at the
origin (big bang) inducing a strain everywhere in space time\footnote{In the case of a spatially flat spacetime we should rather refer to
a singular surface than to a singular event, but the logic structure remains
the same.}. A vector field naturally arises from this view: the "radial"
rate of stress $\gamma $ induced by the defect \cite{dark1}. Now "radial"
means along the cosmic time axis.

The mentioned identification of the vector field implies it to be
divergence-free, which means%
\begin{equation}
\gamma _{;\mu }^{\mu }=0  \label{divenul}
\end{equation}

In the RW symmetry and adopting a co-moving cosmic reference frame, (\ref%
{divenul}) has the solution
\begin{eqnarray}
\gamma ^{0} &=&\frac{Q^{3}}{a^{3}}  \label{soluzione} \\
\gamma ^{i} &=&0  \nonumber
\end{eqnarray}%
where $Q$ is an integration constant and $a$ is the scale factor of the
universe. In fact the time component of the vector is also a measure of its
norm%
\[
\chi ^{2}=\left( \gamma ^{0}\right) ^{2}
\]

This result introduces a first difference with respect to the \AE\ theory
because there the cosmic four-vector is constrained to have unit norm, in
fact coinciding with the four-velocity of an observer co-moving with the
cosmic fluid. This is not the case of the general Lagrangian density (\ref%
{fergri}), where there is no fixed norm constraint.

As for the choice of the Lagrangian density, the CD theory uses another
analogy based on the remark that the phase space of a RW universe is
bidimensional and that it formally coincides with the one describing a point
particle moving across a viscous fluid. Starting from this formal
correspondence the action integral for the only space time is assumed to be
\cite{dark1}:%
\begin{equation}
S=\frac{1}{2c\kappa}\int e^{-g_{\mu \nu }\gamma ^{\mu }\gamma ^{\nu }}\ R\
\sqrt{-g}\ d^{4}x  \label{azione}
\end{equation}%
where $\gamma $ is again the already mentioned four-vector.

In a RW symmetry the action integral, including matter (Friedman Robertson
Walker case), reduces to
\begin{eqnarray}
S &=&S_{g}+S_{m}  \nonumber \\
&=&\mathcal{V}_{k}\left[ -\int \frac{3}{\kappa }e^{-Q^{6}/a^{6}}\left( a^{2}%
\ddot{a}+a\dot{a}^{2}\right) d\tau +\kappa _{0}\int fa^{3}\dot{a}^{2}d\tau
+\varpi \int ha^{3}d\tau \right]   \label{azioneRW}
\end{eqnarray}%
where now $\tau $ is the cosmic time ($=ct$); dots denote cosmic time
derivatives; $\mathcal{V}_{k}$ is the part of the Lagrangian which is not
affected by any variation with respect to the metric; $\kappa _{0}$ and $%
\varpi $ are appropriate  coupling constants; $f$ and $h$ are scalar
functions of $a$ accounting for anything we could widely speaking dub as
"matter".

We remark that the second derivative of $a$ with respect to $\tau $ appears
linearly in the Lagrangian (the integrand of (\ref{azioneRW})). This means
that integrating the corresponding term by parts in the action leads to
\begin{eqnarray*}
\int e^{-Q^{6}/a^{6}}a^{2}\ddot{a}d\tau &=& \\
&\ &\left. e^{-Q^{6}/a^{6}}a^{2}\dot{a}\right\vert _{\tau _{1}}^{\tau
_{2}}-2\int e^{-Q^{6}/a^{6}}\left( 3\frac{Q^{6}}{a^{5}}+a\right) \dot{a}%
^{2}d\tau .
\end{eqnarray*}%
One is then left with a surface term, whose variation is by definition zero,
and a first order derivative term so that in practice the effective
Lagrangian becomes%
\[
\mathcal{L}=\frac{3}{\kappa }e^{-Q^{6}/a^{6}}\left( 6\frac{Q^{6}}{a^{5}}%
+a\right) \dot{a}^{2}+\kappa _{0}fa^{3}\dot{a}^{2}+\varpi ha^{3}.
\]

$\allowbreak $It is now possible to write the Hamiltonian density function
for the system. This is:%
\begin{eqnarray*}
\mathcal{H} &=&\dot{a}\frac{\partial \mathcal{L}}{\partial \dot{a}}-\mathcal{%
L} \\
&=&\left[ \kappa _{0}fa^{3}+\frac{3}{\kappa }e^{-Q^{6}/a^{6}}\left( 6\frac{%
Q^{6}}{a^{5}}+a\right) \right] \dot{a}^{2}-\varpi ha^{3}
\end{eqnarray*}%
and is, as usual, interpreted as the energy density (in the universe). We
may easily verify that%
\begin{equation}
\frac{d\mathcal{H}}{d\tau }=0.  \label{conservazione}
\end{equation}%
We may then write%
\[
\left[ \kappa _{0}fa^{3}+\frac{3}{\kappa }e^{-Q^{6}/a^{6}}\left( 6\frac{Q^{6}%
}{a^{5}}+a\right) \right] \dot{a}^{2}-\varpi ha^{3}=\mathcal{W}=\textit{
constant}.
\]

Then the field equation becomes%
\begin{equation}
\dot{a}^{2}=\frac{\varpi ha^{3}+\mathcal{W}}{\left[ \kappa _{0}fa^{3}+\frac{3%
}{\kappa }e^{-Q^{6}/a^{6}}\left( 6\frac{Q^{6}}{a^{5}}+a\right) \right] }.
\label{dinamica}
\end{equation}%

Actually, if we want to recover the usual meaning of the matter term in a
co-moving reference frame we must choose%
\[
\kappa _{0}=0
\]%
so the rate of expansion equation becomes:

\begin{equation}
\dot{a}^{2}=\frac{\kappa }{3}\frac{\varpi ha^{3}+\mathcal{W}}{%
e^{-Q^{6}/a^{6}}\left( 6\frac{Q^{6}}{a^{5}}+a\right) }.  \label{field}
\end{equation}%
In the absence of a defect, it would be (FRW universe)%
\begin{equation}
\frac{\dot{a}^{2}}{a^{2}}=\frac{8\pi G}{3c^{4}}\rho c^{2}=\frac{\kappa }{3}%
\rho c^{2}  \label{1}
\end{equation}%
Evaluating (\ref{field}) with $Q=\chi =0$, that is looking at the equation
that comes from the action of the CD theory but in the absence of a defect,
we obtain
\begin{equation}
\frac{\dot{a}^{2}}{a^{2}}=\frac{\kappa }{3}\frac{\varpi ha^{3}+\mathcal{W}}{%
a^{3}}.  \label{2}
\end{equation}%
Of course the value of the $\mathcal{W}$ constant depends on the type of
space-time we consider: in the classical empty case (no defect, no matter)
it would be $\mathcal{W}=0$. We then conclude that in order to recover the
classical result, i.e. comparing (\ref{1}) and (\ref{2}), it must be
\begin{eqnarray*}
\varpi  &=&1 \\
h &=&\rho c^{2}
\end{eqnarray*}%
where now $\rho $ represents the usual mass-energy density function.

The final expansion rate equation is

\begin{equation}
\dot{a}^{2}=\frac{\kappa }{3}\frac{\rho c^{2}a^{3}+\mathcal{W}}{%
e^{-Q^{6}/a^{6}}\left( 6\frac{Q^{6}}{a^{5}}+a\right) }  \label{espansione}
\end{equation}

Introducing the new variable $\tilde{a}=a/Q$ and using $Q$ also as the unit
for time ($\tau \rightarrow \tau Q$) we may recast (\ref{espansione}) in the
form
\begin{equation}
\dot{a}^{2}=\frac{\tilde{\kappa}}{3}\frac{\tilde{\rho}c^{2}\tilde{a}^{3}+
\mathcal{W}}{6+\tilde{a}^{6}}\;\tilde{a}^{5}\;e^{1/\tilde{a}^{6}}
\label{espansione1}
\end{equation}%
where the coupling constant $\kappa$, as well as the volume entering the definition of $\rho $%
, have been rescaled on $Q$:

\[
\;\tilde{\kappa}=\frac{\mathfrak{\kappa }}{Q}\;\;\;\;\;;\;\;\;\;\;\;\;\;%
\tilde{\rho}=Q^{3}\rho .
\]%
%
%
%
%
%
%
%
%
%
%
%

\section{The accelerated expansion}

\label{sec4}

\subsection{In the Vector \AE ther theories}

Let us investigate, now, cosmological solutions deduced from the VE\
theories, i.e. from (\ref{ein}). As we already know, in the case of a
homogeneous and isotropic universe the constraints of ZFS and \AE\ theory
force the four vector field to be $U^{\mu }=(1,0,0,0)$. The energy-momentum
tensor for matter, thought of as a perfect fluid, can be written as $%
T_{\alpha \beta }^{m}=\rho c^{2}u_{\alpha }u_{\beta }+p(g_{\alpha \beta
}-u_{\alpha }u_{\beta })$, where $\rho c^{2}$ is the energy density, $p$ is
the pressure and $u^{\mu }$ is the four velocity of the fluid (i.e. $%
g_{\alpha \beta }u^{\alpha }u^{\beta }=1$). Equation (\ref{vector}) for the
vector field can be used to deduce $\lambda $ and put it in the
stress-energy tensor for $U^{\mu }$, eq.(\ref{emt}), so that one is left
with the two Einstein equations:
\begin{eqnarray}
H^{2} &=&\kappa \left(\alpha H^2 \mathcal{F}^{\prime 2}-\frac{1}{6}\mathcal{F}%
M^{2}\right) +\frac{\kappa }{3}\rho c^{2}  \nonumber \\
-H^{2}-2\frac{\ddot{a}}{a} &=&-\kappa \left[ \mathcal{F}^{\prime }\alpha
\left( 2H^{2}+\frac{\ddot{a}}{a}\right) -\dot{\mathcal{F}}^{\prime }\alpha H+%
\frac{1}{2}\mathcal{F}M^{2}\right] +\kappa p,  \label{ein2}
\end{eqnarray}%
where $\alpha $ is a combination of the $c_{i}$'s, namely $\alpha
=c_{1}+3c_{2}+c_{3}$ and $H\equiv \frac{\dot{a}}{a}$. The equations(\ref%
{ein2})\footnote{%
There is a slight difference between the equations written here and those
found in \cite{Zlosnik}, because of the definition of the Lagrangian for the
\AE ther: $\mathcal{L}_{U}$ here is $16\pi G/c^{4}$ times the one found in
the cited article.} govern the evolution of the universe. An accelerated
expansion is indeed obtained choosing, for an appropriate range of $\mathcal{%
K}$ values, \cite{Zlosnik}
\begin{equation}
\mathcal{F}(\mathcal{K})=C(-\mathcal{K})^{n}  \label{f}
\end{equation}%
where $C$ is a constant and $n$ an integer. $\mathcal{K}$ is found to be $3%
\frac{\alpha H^{2}}{M^{2}}$.

Restrictions on the $c_{i}$'s are obtained when studying the consistency of
the theory in the perturbative regime, that is performing classical
perturbations in flat space time, and at a quantum level, when the
Hamiltonian has to be positive definite. The analysis has been performed by
Lim in \cite{Lim} in the case of the \AE\ theory; the Lagrangian for the \AE %
ther turns out to have a scalar-type and a vector-type perturbation (i.e. a
spin-$0$ mode and a spin-$1$ mode). The constraints obtained by Lim are the
following:

\begin{itemize}
\item $c_{1}<0$, in order to insure that spin-$0$ states have positive norm,
i.e. are non-ghost-like;

\item $0\leq \frac{c_{1}+c_{2}+c_{3}}{c_{1}}\leq 1$, in order to have a
well-behaved propagation of the spin-$0$ mode;

\item $c_{1}+c_{3}\geq 0$ in order to have gravitational waves propagating
subluminally.
\end{itemize}

All these conditions together imply
\begin{eqnarray}
c_{1} &<&0  \nonumber \\
c_{2} &\leq &0  \nonumber \\
c_{1}+c_{2}+c_{3} &\leq &0,  \nonumber
\end{eqnarray}%
i.e., $\alpha \leq 0$; this is the reason why $\mathcal{K}$ appears with a
minus sign in $\mathcal{F}$ in (\ref{f}). In \cite{Zlosnik} it is also shown
that one can rewrite the Einstein equations to obtain, with the particular
choice of $\mathcal{F}$ written above in (\ref{f}):
\[
\left[ 1+\epsilon \left( \frac{H}{M}\right) ^{2(n-1)}\right] H^{2}=\frac{%
\kappa }{3}\rho c^2 ,
\]%
where $\epsilon =-(1-2n)C(-3\alpha )^{n}/6$. This solution introduces two
more completely free parameters, besides the ones already present in the
Lagrangian. It is then possible to choose them so that $\epsilon <0$ and
find out that $H$ tends to the attractor: $\tilde{H}=M(-\epsilon
)^{1/2n(1-n)}$. Besides the $c_{i}$'s, $C$ and $n$, the mass scale $M$ is
also present; the authors (ZFS) relate it to the acceleration scale $a_{0}$
of Milgrom's MOND theory \cite{milgrom}, in order to have the right limit at
galactic scale.

Restricting to \AE\ case (cfr. \cite{EE1}) the equations reduce to
\begin{eqnarray}
H^{2} &=&\kappa \;\alpha H^{2}+\frac{\kappa }{3}\rho c^{2}  \nonumber \\
-\left( H^{2}+2\frac{\ddot{a}}{a}\right)  &=&\kappa \left[ -\alpha \left(
H^{2}+2\frac{\ddot{a}}{a}\right) +p\right] .  \label{equazione}
\end{eqnarray}%
The analysis of these equations has been performed by Carroll and Lim in
\cite{Carroll}, but their $\alpha $ parameter is opposite in sign with
respect to the one used here, because we are following the notations of \cite%
{Zlosnik}; furthermore in their case $\kappa \alpha $ is a dimensionless
quantity. \newline
Inspecting the first equation in (\ref{equazione}), which is the $00$-th
component of the Einstein equations, one can easily see that the
contribution from the stress-energy tensor of the vector field is
proportional to the square of the Hubble parameter. In practice the
equations can be rewritten as the usual Friedmann equations just rescaling
the gravitational constant $G$:
\begin{eqnarray*}
H^{2} &=&\frac{\kappa }{3\left( 1-\kappa \alpha \right) }\rho c^{2} \\
\frac{\ddot{a}}{a} &=&-\frac{4\pi G_{c}}{3c^{4}}(\rho c^{2}+3p),
\end{eqnarray*}%
The effective gravitational constant $G_{c}$ is
\begin{equation}
G_{c}\equiv \frac{G}{1-8\pi G\alpha /c^{4}}.  \label{cosmologico}
\end{equation}%
Since $\alpha \leq 0$ the effect of the vector field is to increase the rate
of expansion of the universe, but $G$ is not directly measurable. In order
to obtain constraints on $\alpha $ values, one has to consider other
situations, first of all the Newtonian limit. We note that in \AE\ theory,
i.e. in the last analysis we have outlined, there is no accelerated
expansion, since it retraces the GR solution.


\subsection{In the CD theory}

The interesting feature of eq. (\ref{espansione}) is that it does indeed
contain an accelerated expansion phase in the history of the universe.
Studying the properties of (\ref{espansione}) we see that the expansion rate
starts with an infinite value at the origin and tends to $0$ at infinity.
The initial expansion is exponential, i.e. inflationary;\ at the other end,
for any reasonable behaviour of matter, the expansion continues for ever at
a rate asymptotically tending to $0$.

\begin{figure}[tbp]
\begin{center}
\includegraphics[width=12cm]{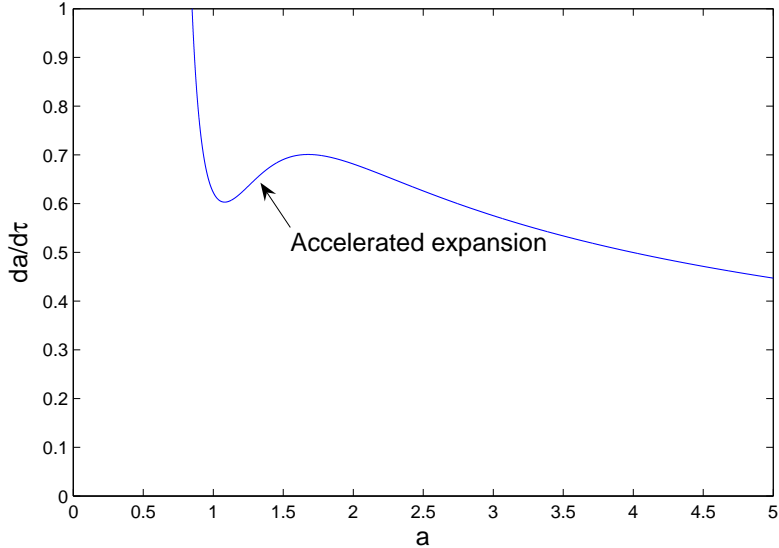}
\end{center}
\caption{Expansion rate of the universe versus the scale factor $a$
according to the CD theory. The graph is valid both for empty space-time and
for a universe filled with an incoherent dust.}
\label{fig:accelerazione}
\end{figure}

If the defect is a property of space-time the expansion (which is our way to
describe what actually is a static state in four dimensions) is present even
in the absence of matter, and, remarkably, one has a sequence of
decelerated-accelerated-decelerated expansion. In fact $\tilde{a}$ from (\ref%
{espansione1}) with $\tilde{\rho}=0$ has two extrema corresponding to%
\begin{equation}
\tilde{a}=\left( 12\pm 6\sqrt{3}\right) ^{1/6}  \label{numeri}
\end{equation}%
The same result is obtained when matter is present in the form of dust. In
that case mass conservation implies%
\[
\rho =\rho _{0}\frac{\tilde{a}_{0}^{3}}{\tilde{a}^{3}}
\]%
leading to a renormalization of constants not modifying (\ref{numeri}). Fig.%
\ref{fig:accelerazione} shows the behaviour of the expansion rate as a
function of the cosmic scale factor.

The situation is different if we allow for more general forms of matter. For
a simple barotropic fluid with an equation of state
\[
\rho c^{2}=wp,
\]%
where $0\leq w\leq 1/3$, the conservation laws of thermodynamics imply that%
\[
\rho =\rho _{0}\frac{\tilde{a}_{0}^{3\left( 1+w\right) }}{\tilde{a}^{3\left(
1+w\right) }}.
\]%
In this case the equation for the extrema, from (\ref{espansione1}), is
\[
\tilde{a}^{3w}\left( 36-24\tilde{a}^{6}+\tilde{a}^{12}\right) \mathfrak{W}-6%
\tilde{a}^{6}\left( 3w -4\right) -\tilde{a}^{12}\left( 1+3w\right) -36=0
\]%

We are left with two free parameters, $w$ and $\mathfrak{W=}\mathcal{W}%
/\left( c^{2}\tilde{\rho}_{0}\tilde{a}_{0}^{3\left( 1+w\right) }\right) $,
to be determined in order to recover both the observed onset of the
accelerated expansion and the age of the universe.


\section{The Newtonian limit}

\label{sec5}

Since General Relativity satisfies all the Solar system tests, any extension
or modification of GR must possess a correct Newtonian limit. In this
section we want to compare the theories we have been discussing until now,
also on this respect. In practice we expect that, given any material source,
the field equations for gravity, in weak field approximation, reduce to the
Poisson equation for the potential.

\subsection{In Vector \AE ther Theories}

Let us consider the field equations (\ref{ein}) in the static, weak field
limit. The way chosen both in \cite{Zlosnik} and in \cite{Carroll} is to
expand both the metric and the vector field around a Minkowski background.
At the lowest non-trivial order the approximated line element may be written
as follows:
\[
ds^{2}=(1+2\Phi (x,y,z))d\tau ^{2}-(1-2\Psi (x,y,z))(dx^{2}+dy^{2}+dz^{2}),
\]%
where $\Phi $ and $\Psi $ are suitable potentials. Since we are in the weak
field limit, we shall neglect terms beyond the first order in the
potentials; under this assumption the space components of the \AE ther
stress-energy tensor disappear. The spatial components of the Einstein
equations reduce then to
\begin{equation}
(\delta _{ij}\nabla ^{2}-\partial _{i}\partial _{j})(\Phi \;-\;\Psi )=0,
\label{condizione}
\end{equation}%
where $i$ and $j$ are space indices ranging from $1$ to $3$. Assuming that
both $\Psi $ and $\Phi $ vanish at space infinity, (\ref{condizione})
implies that $\Psi =\Phi $. Using this result, while combining the
linearized $00$-th component of the Einstein equations and the vector field
equation, we obtain
\begin{equation}
\vec{\nabla}\cdot \left[ (2+\frac{16\pi G}{c^{4}}c_{1}\mathcal{F}^{\prime })%
\vec{\nabla}\Phi \right] =\frac{8\pi G}{c^{2}}\rho ,  \label{eterpoisson}
\end{equation}%
where $\rho $ is the mass density of the matter distribution. In the
framework of \AE\ theory, it is simply $\mathcal{F}^{\prime }=1$ and
equation (\ref{eterpoisson}) becomes
\[
\nabla ^{2}\Phi (1+\frac{8\pi G}{c^{4}}c_{1})=\frac{4\pi G}{c^{2}}\rho ,
\]%
which is indeed a Poisson equation, with an effective gravitational constant
\begin{equation}
G_{N}=\frac{G}{1+c_{1}\frac{8\pi G}{c^{4}}}.  \label{solare}
\end{equation}%
The results (\ref{solare}) and (\ref{cosmologico}) can be used to obtain one
more constraint on the parameters of the theory, as analyzed in \cite%
{Carroll} and \cite{Lim}.

In the ZFS version of the theory \cite{Zlosnik}\footnote{%
Note that, again, the definition of the Lagrangian for the \AE ther $%
\mathcal{L}_{U}$ here is $\frac{16\pi G}{c^{4}}$ times the one found in the
cited article.}, the authors are first led to identify the mass scale $M$
with something of the order of $a_{0}$, as we have seen before, in order to
recover the MOND limit of the theory. For them it actually is
\[
\mathcal{K}=-c_{1}\frac{(\vec{\nabla}\Phi )^{2}}{M^{2}}
\]%
and $c_{1}<0$ in order to avoid ghost-like spin-$0$ states (see \cite{Lim}).
Now, looking at (\ref{eterpoisson}), we see that, in order to recover, at
least at the Solar system scale, the Poisson equation, the $\mathcal{F}%
^{\prime }$ contribution must be small. For these reasons, the authors
assume that in the Solar system $\left( \vec{\nabla}\Phi \right) ^{2}\gg
M^{2}$ and expand $\mathcal{F}^{\prime }$ as a series of inverse powers of $%
\mathcal{K}^{1/2}$. At this point, however, the non-linearity due to $%
\mathcal{F}^{\prime }$ makes the equations very difficult, thus making very
hard to draw clear-cut conclusions, as remarked in \cite{bekenstein}.


\subsection{In the CD theory}

In the case of the CD theory, as in the previous section, we have to
consider the weak-field limit of the theory, expanding both the metric and
the vector field around a background configuration, but now we choose the
FRW rather than Minkowski spacetime, because we want to maintain a link
between the cosmological and the local solution. The source of the
perturbation is assumed to be some local, static matter distribution,
superposed to the cosmic one. The details of the whole procedure may be
found in \cite{dark1}; the essentials are outlined in the following.

The perturbed line element is now written as:
\begin{equation}
ds^{2}=(1+h_{0}(x,y,z))c^{2}dt^{2}-a^{2}(t)(1+h_{s}(x,y,z))(dx^{2}+dy^{2}+dz^{2}),
\label{metricapert}
\end{equation}%
with $h_{0},h_{s}<<1$.

We expect the flow lines of the cosmic vector field to be perturbed as well,
however preserving the norm of the vector, which depends on the presence of
the cosmic defect only. Let us write the perturbed vector as $\Upsilon $;
its components (first order approximation) will be:
\begin{eqnarray}
\Upsilon ^{0} &=&\chi (1+f^{0}(x,y,z))  \nonumber \\
\Upsilon ^{i} &=&\chi f_{s}^{i}(x,y,z),  \label{le upsilon}
\end{eqnarray}%
The perturbations are scaled on the unperturbed norm $\chi $, and we assume
that $f^{0}$, $f_{s}^{i}<<1$ (at least as small as the $h$'s), and depend on
the space coordinates. The time dependence is contained in the scale factor $%
a(t)$ only. The divergencelessness condition (\ref{divenul}) applied to $%
\Upsilon $ must still hold, because it is broken only at the site of a
space-time defect and in this respect nothing has changed, in the sense that
no other singularities have been introduced, besides the cosmic one. Eq. (%
\ref{divenul}), at first order in the perturbations, becomes
\begin{equation}
\vec{\nabla}\cdot \vec{f}_{s}=0.  \label{nullpert}
\end{equation}%
The invariance of the norm of $\Upsilon $ produces the condition
\begin{equation}
f^{0}=-\frac{h_{0}}{2}.  \label{fnothnot}
\end{equation}

The next steps may be summarized as follows: a) introduce the metric (\ref%
{metricapert}) in the CD action integral (\ref{azione}) plus matter, then
linearize it in the perturbations; b) deduce the field equations for the
geometry (the equivalent of the Einstein equations); c) consider that the
zero order of the equations is automatically satisfied with the cosmic fluid
energy momentum tensor; d) write down the first order equations with the
local matter energy momentum tensor $\mathcal{T}_{\mu \nu }$ assumed to be
isotropic in space around any given point. You will get:
\begin{equation}
-e^{-\chi^2}\left[\nabla ^{2}h_{s}+\chi ^{2}(\nabla ^{2}h_{0}+2\nabla ^{2}h_{s})\right]=\frac{4\pi G}{%
c^{4}}\mathcal{T}_{00},  \label{equpertu}
\end{equation}%
where $\nabla ^{2}=\frac{1}{a^{2}}(\partial _{x}^{2}+\partial
_{y}^{2}+\partial _{z}^{2})$ and $\mathcal{T}_{00}$ is the energy density of
the local source. As we know, there is a freedom for the choice of the
coordinates, so that the Lorentz gauge can be imposed, leading to $%
h_{s}=-h_{0}$. The final equation is then
\begin{equation}
\nabla ^{2}h_{0}=\frac{4\pi }{c^{4}}\frac{Ge^{\chi ^{2}}}{1+\chi ^{2}}%
\mathcal{T}_{00}.  \label{finalissima}
\end{equation}%
This equation is the Poisson equation with a renormalized gravitational
"constant" slowly changing with time:
\begin{equation}
G_{\ast }=\frac{Ge^{\chi ^{2}}}{1+\chi ^{2}}.  \label{gstar}
\end{equation}%
The cosmic vector field $\gamma $ does indeed affect the local gravitational
field, through its norm $\chi $. This influence is not perceivable on the
usual time scales, in the sense that the Newtonian behaviour is fully
recovered; however in cosmic times the effective coupling "constant" of
gravity, in the Newtonian formalism, slowly changes. Had we started from a
Minkowski background, this adiabatic effect would not have been visible, as
it happens in the \AE\ theory where two formally different renormalizations
of $G$ are obtained at the cosmic and at the local scale.


\section{Correspondence between the theories}

\label{sec6}

VET, as well as the more general SME, and CD are apparently rather
different from each other, however, as we shall show here, it is possible to
recast the latter in a form which will make it emerge as a special case of
the former. The comparison will then be made at the level
of the effective action integrals.

Considering the CD action (\ref{azione}) we remark that it could be thought
of as being the result of a conformal transformation from some previous
appropriate metric. To evidence this interpretation in the present section,
we shall mark the entities used in the CD theory with a $\sim $ assuming
that $\tilde{g}_{\mu \nu }=e^{2\omega }g_{\mu \nu }$, being $\omega $ a
conformal factor. Let us rewrite (\ref{azione}) accordingly:
\begin{equation}
S=\frac{1}{2\kappa c}\int e^{-\chi ^{2}}\tilde{R}\ \sqrt{-\tilde{g}}\ d^{4}x.
\label{azioneconforme}
\end{equation}%
Consistently with the approach we are describing here, the curvature and the
square root of the determinant of the metric tensor may be written as%
\begin{eqnarray}
\tilde{R} &=&e^{-2\omega }\left[ R-6g^{\mu \nu }\nabla _{\mu }\nabla _{\nu
}\omega -6g^{\mu \nu }(\nabla _{\mu }\omega )\left( \nabla _{\nu }\omega
\right) \right]  \label{R} \\
\sqrt{-\tilde{g}} &=&e^{4\omega }\sqrt{-g}\\
\tilde{\chi}^2&=&e^{2\omega}\chi^2\label{detconf}
\end{eqnarray}

If we now choose the conformal factor so that
\begin{equation}\label{trasc}
\chi^2=2\omega e^{-2\omega} ,
\end{equation}
since of course $ e^{2\omega}e^{-\chi^2 e^{2\omega}}=1 $, the effective Lagrangian
density before the transformation turns out to be%
\begin{equation}
\left[ R-6g^{\mu \nu }\nabla _{\mu }\nabla _{\nu }\omega -6g^{\mu \nu
}(\nabla _{\mu }\omega )\left( \nabla _{\nu }\omega \right) \right] \sqrt{-g}
\label{elle}
\end{equation}%
Recalling that $\chi ^{2}$ is the norm of the vector field $\gamma ^{\mu }$,
we can rewrite the second and third terms in the square brackets of (\ref%
{elle}) as explicitly depending on $\gamma $. 

The solution to this trascendental equation is the Lambert function, in particular
$$
2\omega=-W_k(-\chi^2).
$$

The Lambert function $W_k(z)$ is a multivalued function of the complex variable $z$ and $k$ is an integer that represents the branch we are looking at. In our case the variable is the norm of the vector field that is time-like: we must restrict to the case in which $z$ is real, let us say $x$, and $x>0$. Furthermore, we want the conformal factor, as well as the Lambert function, to be real, so that we consider only the case in which the argument is greater than $-1/e$. But in our solution the argument of the Lambert function is $-\chi^2$, that is, the norm of the vector filed can change only between $0$ and $1/e$. In order to enhance this range we can solve the subsequent equation, instead of \ref{trasc}:

\begin{equation}
 e^{2\omega}e^{-\chi^2 e^{2\omega}}=const \Rightarrow \chi^2e^{2\omega}-2\omega=c.
\end{equation}

This simply means that we are changing the value of the dimensional constant in front of the action integral. The solution, now, is

$$
2\omega=c-W_k(-e^c \chi^2)
$$
and the range of the variable for which the Lambert function is real. To enhance it we have to choose $c<1$. 

But let us note that in the range we are interested in the Lambert function is two-valued. For $W(x)\geq-1$  the function is denoted $W_0(x)$, or simply $W(x)$, and is called the principal branch; for $W(x)\leq-1$ the function is denoted $W_{-1}(x)$.

If we look for the principal branch, a Taylor series can be found but due to the singularity at $x=-1/e$ the series converges for $|x|<1/e$. The series is

$$
W(x)=\sum_{n=1}^{\infty}\frac{(-n)^{n-1}}{n!}x^n=x-x^2+\frac{3}{2}x^3+\dots
$$

Rewriting the solution in terms of this series, and remebering that the action integral of VET is fourth order in the fields, we can drop all terms beyond second order, since the variable $x$ corresponds to the norm of the vector field, that is second order in the field.

Now, we are left only with the terms that appear in $\tilde{R}$. Apart from the first term, that is the Ricci scalar in terms of the old metric, there are two additional terms. Let us look at the first:
\begin{eqnarray}
g^{ab}\nabla_a\nabla_b \omega &\sim&g^{ab}\nabla_a\nabla_b (c+e^c\chi^2-e^{2c}\chi^2\chi^2)\nonumber\\
&=&2e^cg^{ab}\nabla_a\left(\gamma^c\nabla_b \gamma_c\right)-e^{2c}g^{ab}\nabla_a\nabla_b(\chi^2 \chi^2)\nonumber\\
&=&2e^cg^{ab}\nabla_a\left(\gamma^c\nabla_b \gamma_c\right)-2e^{2c} g^{ab} \nabla_a(\chi^2\nabla_b \chi^2)
\end{eqnarray}
where both terms are divergence, that is they reduce to surface terms when integrating. We are left only with the second term in (\ref{R}), in which, in order to obtain terms up to fourth order, reduce to the following one:
\begin{eqnarray}
g^{ab}(\nabla_a \omega)(\nabla_b\omega)&\sim&g^{ab}\nabla_a(e^c \chi^2)\nabla_b (e^c \chi^2)\nonumber\\
&=&e^{2c}g^{ab}\left(2 \gamma^c\nabla_a \gamma_c\right)\left(2 \gamma^d\nabla_b \gamma_d\right)=4e^{2c} g^{ab} \gamma^c \gamma^d \nabla_a \gamma_c \nabla_b \gamma_d.
\end{eqnarray}

Our effective Lagrangian density is now:%
\begin{equation}
\left( R-6e^{2c}g^{\mu \nu }\gamma ^{\alpha }\gamma ^{\beta }\nabla _{\mu }\gamma
_{\alpha }\nabla _{\nu }\gamma _{\beta }\right) \sqrt{-g}
\label{daconfronto}
\end{equation}

What here is called $\gamma $, in the VE \ theories is the $U$ vector, so
that comparing (\ref{daconfronto}) with the Lagrangian density in (\ref%
{fergri}) we see that the CD theory corresponds to a VE\ theory with the
only $c_{6}$ coefficient differing from $0$. Actually it is $c_{6}=-6e^{2c}$.


As we wrote in section (\ref{sec2}), VET may be considered as
special cases of the SME theory so that a direct comparison to that theory
is appropriate. A simple example of a cosmic vector field is the so
called "bumblebee" vector field illustrated in appendix\ B\ of ref.\cite{kos}%
. The bumblebee field is indeed a timelike vector $B$
dynamically depending on a suitable potential in a torsionless
spacetime. Writing $B_{\mu \nu }=\left( \nabla _{\mu }B_{\nu }-\nabla _{\nu
}B_{\mu }\right) $, the Lagrangian density is assumed to be
\begin{equation}
\mathcal{L}_{B}=\frac{\xi }{2\kappa }B^{\mu }B^{\nu }R_{\mu \nu }-\frac{1}{4}%
B^{\mu \nu }B_{\mu \nu }-V\left( B^{\mu }B_{\mu }\pm b^{2}\right)
\label{bumble}
\end{equation}%
where $\xi $ is a parameter and $V\left( x\right) $
is a scalar potential; $b$ sets the position of the minimum
of the potential.  Comparing (\ref{bumble}) with (\ref{fergri}) and (\ref%
{Ikappa}) we see that the former corresponds to the latter for peculiar
combinations of the $c_{i}$'s (of course to see this one must
express $R_{\mu \nu }$ in terms of the $g_{\mu \nu }$'s
and their derivatives, then making some integrals by parts in the action).
In particular the  "bumblebee" model is recovered when %
$ c_{1}=-1/2,c_{3}=1/2-c_{2}$.

\section{$\allowbreak $Conclusion}

We have been analyzing in parallel the CD theory, on one side, and the VE\
theories (and especially the Einstein \AE ther theory) on the other. Both
(groups of) theories are based on the presence of a cosmic timelike vector
field, and may be thought of as special cases of the SME theory.

The VET\ contain a big number of adjustable parameters, and, in the ZFS\
case, also a free function. Various ways to exploit this wide freedom allow
for different approaches and lead to different conclusions. In fact, rather
than a global scenario, a number of specific, not completely consistent
pictures emerge. For instance the accelerated expansion is present in one
version, and not in others; the gravitational coupling constant has formally
different limits in the cosmic and in the Newtonian limit. Furthermore the
physical meaning of the majority of the free parameters of the theory is
unclear.

In the CD theory only one free parameter exists in the description of space
time: a sort of global scale constant, to be determined on the basis of the
observed behaviour of the universe. One more parameter appears when
considering the coupling between matter and spacetime. The theory accounts
for the accelerated expansion and possesses a Newtonian limit with a
renormalized gravitational "constant" slowly changing in time.

Apart from the details, an important difference between VET\ and CD is in
the embedding paradigm. In the case of VET\ we are in the mathematical
framework of vector-tensor field theories, and the hypothesized vector
field, as well as many parameters, lack a physical motivation other than the
final result. In the case of CD the paradigm is based on some analogies with
problems of the physics we already know, and the vector field is thought of
as the consequence of the strain induced in a four-dimensional medium
(spacetime) by the presence of a defect, in the sense of the elasticity
theory. This paradigm makes the theory very compact, minimizing the number
of adjustable coefficients and making the comparison with observation easier
or, at least, the conclusions sharper.

The fact that the real main difference is in the interpretation paradigms
appears clearly when, as we did in Sec. (\ref{sec6}), we show that CD may be
looked at as to a special case of VET. However, had we gone from the Vector
\AE ther theory to the Cosmic Defect, the corresponding peculiar choice of
the parameters of VET\ would have appeared to be completely arbitrary. On
the contrary the approach used for CD provides a consistent interpretation
scheme, which in the end is shown to be mathematically equivalent to one
specific \AE ther theory.

For these reasons we think that the CD paradigm can be fruitfully exploited
again for a deeper understanding of the evolution of our universe.

\label{conc}


\end{document}